\documentclass[twocolumn,showpacs,preprintnumbers,amsmath,amssymb]{revtex4}
\usepackage{graphicx}
\usepackage[ansinew]{inputenc}

\begin{document}

\bibliographystyle{naturemag}

\title{State-independent experimental test of quantum contextuality}

\author{G. Kirchmair$^{1,2}$}
\author{F. Z\"ahringer$^{1,2}$}
\author{R. Gerritsma$^{1,2}$}
\author{M. Kleinmann$^{1}$}
\author{O. G\"uhne$^{1,3}$}
\author{A. Cabello$^{4}$}
\author{R. Blatt$^{1,2}$}
\author{C. F. Roos$^{1,2}$}
\email{christian.roos@uibk.ac.at}

\affiliation{$^1$Institut f\"ur Quantenoptik und Quanteninformation,
\"Osterreichische Akademie der Wissenschaften, Otto-Hittmair-Platz
1, A-6020 Innsbruck, Austria\\
$^2$Institut f\"ur Experimentalphysik, Universit\"at Innsbruck, Technikerstr.~25, A-6020 Innsbruck, Austria\\
$^3$Institut f{\"u}r theoretische Physik, Universit{\"a}t
Innsbruck, Technikerstr.~25, A-6020 Innsbruck, Austria\\
$^4$Departamento de F\'isica Aplicada II, Universidad de Sevilla, E-41012 Sevilla, Spain}

\date{\today}

\begin{abstract}
The question of whether quantum phenomena can be explained by
classical models with hidden variables is the subject of a long
lasting debate\cite{Einstein35}. In 1964, Bell showed that certain
types of classical models cannot explain the quantum mechanical
predictions for specific states of distant particles\cite{Bell64}.
Along this line, some types of hidden variable models have
been experimentally ruled out\cite{Aspect82, Tittel98, Weihs98,
Rowe01, Groeblacher07, Branciard08,  Matsukevich08}. An intuitive
feature for classical models is non-contextuality: the property that
any measurement has a value which is independent of other compatible
measurements being carried out at the same time. However,
the results of Kochen, Specker, and Bell\cite{Specker60,
KochenSpecker67, Bell66} show that non-contextuality is in conflict
with quantum mechanics. The conflict resides in the
structure of the theory and is independent of the properties of special states. It has
been debated whether the Kochen-Specker theorem could be
experimentally tested at all\cite{Cabello98, Meyer99}. Only
recently, first tests of quantum contextuality have been proposed and undertaken
with photons\cite{Huang03} and neutrons\cite{Hasegawa06,Bartosik09}. Yet
these tests required the generation of special quantum states and
left various loopholes open. Here, using trapped ions, we
experimentally demonstrate a state-independent conflict with
non-contextuality. The experiment is not subject to the detection
loophole and we show that, despite imperfections and possible
measurement disturbances, our results cannot be explained in
non-contextual terms.
\end{abstract}


\maketitle

Hidden variable models assert that the result $v(A)$ of
measuring the observable $A$ on an individual quantum system
is predetermined by a hidden variable $\lambda$. Two observables
$A$ and $B$ are mutually compatible, if the result of $A$
does not depend on whether $B$ is measured before, after, or
simultaneously with $A$ and vice versa. Non-contextuality is
the property of a hidden variable model that the value $v(A)$ is determined,
regardless of which other compatible observable is measured jointly
with $A$. As a consequence, for compatible observables the relation
$v(AB)=v(A)v(B)$ holds. Kochen and Specker showed that the
assumption of non-contextuality cannot be reconciled with quantum
mechanics. A considerable simplification of the original
Kochen-Specker argument by Mermin and Peres\cite{Peres90,
Mermin90} uses a $3 \times 3$ square of observables $A_{ij}$ with
possible outcomes $v(A_{ij})=\pm 1$, where the observables in each row or column are mutually compatible. Considering the products of
rows $R_k=v(A_{k1})v(A_{k2})v(A_{k3})$ and columns
$C_k=v(A_{1k})v(A_{2k})v(A_{3k})$, the total product would be
$\prod_{k=1,2,3} R_k C_k =1$, since any $v(A_{ij})$ appears twice in
the total product.

In quantum mechanics, however, one can take a four-level quantum system,
for instance two spin-$\frac{1}{2}$-particles, and the following
array of observables,
\begin{equation}
\begin{array}{ccc}
A_{11}=\sigma_z^{(1)} &A_{12}= \sigma_z^{(2)} &A_{13}= \sigma_z^{(1)}\otimes\sigma_z^{(2)}\\
A_{21}=\sigma_x^{(2)} &A_{22}= \sigma_x^{(1)} &A_{23}= \sigma_x^{(1)}\otimes\sigma_x^{(2)}\\
A_{31}=\sigma_z^{(1)}\otimes\sigma_x^{(2)} &A_{32}= \sigma_x^{(1)}\otimes\sigma_z^{(2)} &
A_{33}=\sigma_y^{(1)}\otimes\sigma_y^{(2)}.
\end{array}
\label{eq1:MPSquare}
\end{equation}
Here, $\sigma^{(k)}_i$ denotes the Pauli matrix acting on the
$k$-th particle, and all the observables have the outcomes
$\pm 1$. Moreover, in each of the rows or columns of
(\ref{eq1:MPSquare}), the observables are mutually commuting and can be measured simultaneously or in any
order. In any row or column, their measurement product $R_k$ or
$C_k$ equals 1, except for the third column where it equals $-1$. Hence, quantum mechanics yields for
the product $\prod_{k=1,2,3} R_kC_k$ a value of $-1$, in contrast
to non-contextual models.

To test this property, it has to be expressed as an inequality since no experiment
yields ideal quantum measurements. Recently, it has been shown that the inequality
\begin{equation}
\label{eq2:KSInequality}
\langle \mathcal{X}_{\rm KS} \rangle=\langle R_1\rangle+\langle R_2\rangle+\langle R_3\rangle+\langle C_1\rangle+\langle C_2\rangle-\langle C_3\rangle\le 4
\end{equation}
holds for all non-contextual theories\cite{Cabello08}, where $\langle\cdots\rangle$ denotes the ensemble average. Quantum mechanics predicts for {\it any} state that $\langle \mathcal{X}_{\rm KS} \rangle=6$, thereby violating inequality (\ref{eq2:KSInequality}). For an experimental test, an ensemble of
quantum states $\Psi$ needs to be prepared and each realization
subjected to the measurement of one of the possible sets of compatible
observables. Here, it is of utmost importance that all measurements
of $A_{ij}$ are context-independent\cite{Cabello08}, i.e., $A_{ij}$ must be
detected with a quantum non-demolition (QND) measurement that provides no
information whatsoever about any other co-measurable observable.

\begin{figure}[t]
\includegraphics[width=8cm]{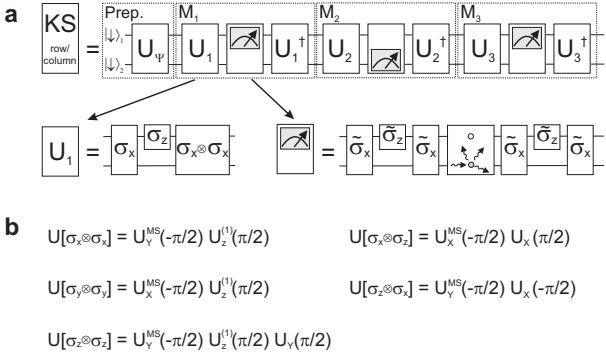}
\caption{\label{fig1:ExpSequence} Experimental measurement scheme. {\bf a} For the measurement of the $j$th row (column) of the Mermin-Peres square (\ref{eq1:MPSquare}), a quantum state is prepared on which three consecutive QND measurements $M_k$, $k=1,2,3$, are performed measuring the observables $A_{jk}$ ($A_{kj}$). Each measurement consists of a composite unitary operation $U_k$ that maps the observable of interest onto one of the single-qubit observables $\sigma_z^{(1)}$ or $\sigma_z^{(2)}$ which are measured by fluorescence detection. To this end, the quantum state of the qubit that is not to be detected is hidden\cite{Roos04} in the $D_{5/2}$-Zeeman state manifold by a composite $\pi$-pulse transferring the $|\!\downarrow\rangle$ state's population to the auxiliary state $|a\rangle\equiv|D_{5/2},m=5/2\rangle$ prior to fluorescence detection. After the detection, the qubit state is restored. In the lower line, $\sigma_i$, $\tilde{\sigma}_i$ symbolize the Hamiltonian acting on the qubit or on the subspace spanned by $\{|\!\downarrow\rangle,|a\rangle\}$. The unitary operations $U$ are synthesized from single-qubit and maximally entangling gates. {\bf b} All mapping operations $U_k$ employed for measuring the five two-qubit spin correlations $\sigma_i^{(1)}\otimes\sigma_j^{(2)}$ require an entangling gate. Here, we list the gate decompositions of $U[\sigma_i^{(1)}\otimes\sigma_j^{(2)}]$ used in the experiments where $U_X(\theta)\equiv U(\theta,\phi=0)$ and $U_Y(\theta)\equiv U(\theta,\phi=\pi/2)$.}
\end{figure}

Experiments processing quantum information with trapped
ions\cite{Haffner08} are particularly well-suited for this
purpose as arbitrary two-qubit quantum states $\Psi$ can be
deterministically generated by laser-ion interactions and measured
with near-unit efficiency\cite{Rowe01}. Two ionic
energy levels are defined to represent the qubit basis states
$|\!\uparrow\rangle$ and $|\!\downarrow\rangle$, which are eigenstates
of the observable $\sigma_z$. The qubit is measured by electron
shelving\cite{Haffner08} projecting onto $|\!\uparrow\rangle$ or
$|\!\downarrow\rangle$. Measurement of any other observable $A_{ij}$
is reduced to detecting $\sigma_z^{(k)}$, k=1 or 2, by applying
a suitable unitary transformation $U$ to the state $\Psi$ prior to
measuring $\sigma_z^{(k)}$, and its inverse $U^\dagger$ after
the measurement (see Fig.~\ref{fig1:ExpSequence} and Methods). With
these basic tools, any set of observables can be sequentially
measured in an arbitrary temporal order.

For the experiment, a pair of $^{40}$Ca$^+$ ions is trapped in a linear Paul trap with axial and radial vibrational frequencies of $\omega_{ax}=(2\pi)\,1.465$~MHz and $\omega_r\approx(2\pi)\,3.4$~MHz and Doppler-cooled by exciting the $S_{1/2}\leftrightarrow P_{1/2}$ and $P_{1/2}\leftrightarrow D_{3/2}$ dipole transitions. Optical pumping initializes an ion with a fidelity of $99.5\%$ to the qubit state $|\!\downarrow\rangle\equiv |S_{1/2},m=1/2\rangle$, the second qubit state being $|\!\uparrow\rangle\equiv |D_{5/2},m=3/2\rangle$. The qubit is coherently manipulated\cite{Kirchmair09} by an ultrastable, narrowband laser coherently exciting the $S_{1/2}\leftrightarrow D_{5/2}$ quadrupole transition in a magnetic field of $B=4$ Gauss.
Single-qubit light-shift gates $U_z^{(1)}(\theta)=\exp(-i\frac{\theta}{2}\sigma_z^{(1)})$ are realized by an off-resonant beam impinging on ion 1 with a beam waist of $3\,\mu$m and a k-vector perpendicular to the ion string. A second beam, illuminating both ions with equal strength at an angle of 45$^\circ$ with respect to the ion crystal, serves to carry out gate operations that are symmetric under qubit exchange. Collective single-qubit gates $U(\theta,\phi)=\exp(-i\frac{\theta}{2}(\sigma_\phi^{(1)}+\sigma_\phi^{(2)}))$, where $\sigma_\phi=\cos(\phi)\sigma_x+\sin(\phi)\sigma_y$,
are realized by resonantly exciting the qubit transition and controlling the phase $\phi$ of the laser light. If instead a bichromatic light field near-resonant with the upper and lower sideband transitions of the axial
centre-of-mass (COM)
mode is used, a M{\o}lmer-S{\o}rensen gate\cite{Molmer99} $U^{MS}(\theta,\phi)=\exp(-i\frac{\theta}{2}\sigma_\phi^{(1)}\otimes\sigma_\phi^{(2)})$ is implemented\cite{Benhelm08,Kirchmair09}.
We achieve a maximally entangling gate ($\theta=\pi/2$) capable of mapping $|\!\downarrow\downarrow\rangle$ to $|\!\downarrow\downarrow\rangle +ie^{-i2\phi}|\!\uparrow\uparrow\rangle$ with a fidelity of about 98\% even for Doppler-cooled ions in a thermal state with
an average of $\bar{n}_{ax,COM}\approx 18$ vibrational quanta. This property is of crucial importance as the experiment demands gate operations subsequent to quantum state detection by
fluorescence measurements which do not preserve the motional quantum
number. The set of elementary gates\cite{Nebendahl09}
$\{U_z^{(1)}(\theta),U(\theta,\phi),U^{MS}(\theta,\phi)\}$ is
sufficient to construct the two-qubit unitary operations needed for
creating various input states $\Psi$ and mapping the observables
$A_{ij}$ to $\sigma_z^{(k)}$ for read-out (see Methods).
\begin{figure}
\includegraphics[width=8cm]{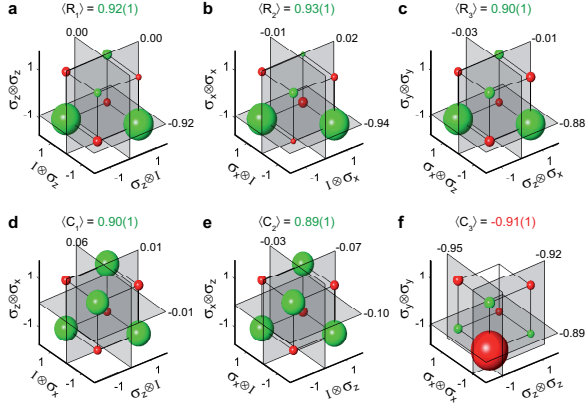}
\caption{\label{fig2:MeasurementCorrelations} Measurement correlations for the singlet state. {\bf a} This subplot visualizes the consecutive measurement of the three observables $A_{11}=\sigma_z^{(1)}$, $A_{12}=\sigma_z^{(2)}$, $A_{13}=\sigma_z^{(1)}\otimes\sigma_z^{(2)}$ corresponding to row 1 of the Mermin-Peres square. The measurement is carried out on 1,100 preparations of the singlet state. The volume of the spheres on each corner of the cube represents the relative frequency of finding the measurement outcome $\{v_1,v_2,v_3\},\,v_i\in\{\pm 1\}$. The color of the sphere indicates whether $v_1v_2v_3=+1$ (green) or $-1$ (red). The measured expectation values of the observables $A_{1j}$ are indicated by the intersections of the shaded planes with the axes of the coordinate system. The average of the measurement product $\langle R_1\rangle$ is given at the top. {\bf b-f} Similarly, the other five subplots represent measurements of the remaining rows or columns of the Mermin-Peres square. Subplot {\bf f} demonstrates that the singlet state is a common eigenstate of the observables $\sigma_x^{(1)}\otimes\sigma_x^{(2)}$, $\sigma_y^{(1)}\otimes\sigma_y^{(2)}$, $\sigma_z^{(1)}\otimes\sigma_z^{(2)}$, as only one of the spheres has a considerable volume. Taking into account all the results, we find $\langle \mathcal{X}_{KS} \rangle= 5.46(4)$ in this measurement. Error bars, $1\sigma$.}
\end{figure}

Equipped with these tools, we create the singlet state $\Psi=(|\!\uparrow\downarrow\rangle-|\!\downarrow\uparrow\rangle)/\sqrt{2}$ by applying the gates
$U_z^{(1)}(\pi)U(\frac{\pi}{2},\frac{3\pi}{4})U^{MS}(\frac{\pi}{2},0)$
to the initial state $|\!\downarrow\downarrow\rangle$ and measure consecutively the three observables of a row or column of the Mermin-Peres square. The results obtained for a total of 6,600 copies of $\Psi$ are visualized in Fig.~\ref{fig2:MeasurementCorrelations}. The three upper panels show the distribution of measurement results
$\{v(A_{i1}),v(A_{i2}),v(A_{i3})\}$, their products
as well as the expectation values $\langle A_{ij}\rangle$ for the observables appearing in the rows of (\ref{eq1:MPSquare}), the three lower panels show the corresponding results for the columns of the square. All of the correlations have a value close to $+1$ whereas $\langle C_3\rangle=-0.913$. By adding them up and subtracting $\langle C_3\rangle$, we find a value of $\langle \mathcal{X}_{KS} \rangle=5.46(4)>4$, thus violating equation~(\ref{eq2:KSInequality}).
\begin{figure}
\includegraphics[width=8cm]{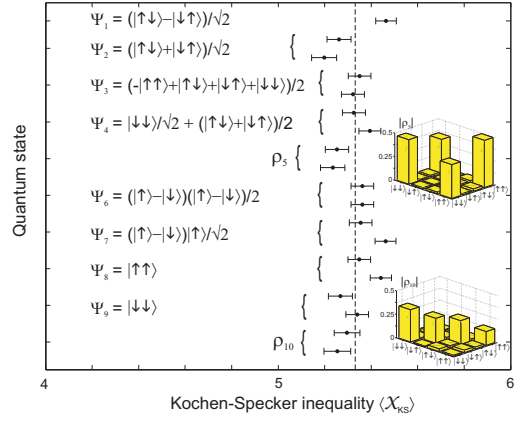}
\caption{\label{fig3:AllStates} State-independence of the Kochen-Specker inequality. The Kochen-Specker inequality was tested for ten different quantum states, including maximally entangled ($\Psi_1$-$\Psi_3$), partially entangled ($\Psi_{4}$) and separable ($\Psi_6$-$\Psi_9$) almost pure states as well as an entangled mixed state $(\rho_{5}$) and an almost completely mixed state ($\rho_{10}$). All states are analysed by quantum state tomography which yields for the experimentally produced states $\Psi_1$-$\Psi_4$, $\Psi_6$-$\Psi_9$ an average fidelity of 97(2)\%. For all states, we obtain a violation of inequality~(\ref{eq2:KSInequality}) which demonstrates its state-independent character, the dashed line indicating the average value of $\langle{\mathcal X}_{KS}\rangle$. Error bars, $1\sigma$ (6,600 state realizations per data point).}
\end{figure}

\begin{figure}[t]
\includegraphics[width=8cm]{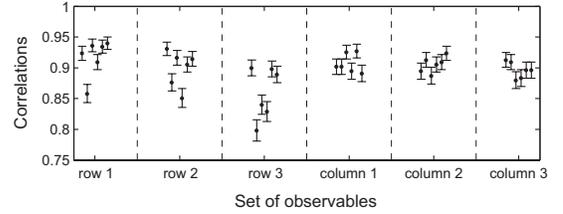}
\caption{\label{fig4:PermutatedTriples} Permutation within rows and columns of
the Mermin-Peres square. As the three observables of a set are commuting,
the temporal order of their measurements should have no influence on the
measurement results. The figure shows the measured absolute values of the products of observables for any of the six possible permutations. By combining the results for the measurement of rows and columns, we obtain 36 values for the Kochen-Specker inequality ranging from 5.22 to 5.49, the average value given by 5.38. The scatter in the experimental data is caused by experimental imperfections that affect different permutations differently. For the measurements shown here, in total 39,600 copies of the singlet state were used.}
\end{figure}

To test the prediction of a state-independent violation, we repeated the experiment for nine other quantum states of different purity and entanglement. Figure~\ref{fig3:AllStates} shows that indeed a state-independent violation of the Kochen-Specker inequality occurs, $\langle \mathcal{X}_{KS} \rangle$ ranging from 5.23(5) to 5.46(4). We also checked that a violation of (\ref{eq2:KSInequality}) occurs irrespective of the temporal order of the measurement triples. Figure~\ref{fig4:PermutatedTriples} shows the results for all possible permutations of the rows and columns of (\ref{eq1:MPSquare}) based on 39,600 realizations of the singlet state. When combining the correlation results for the 36 possible permutations of operator orderings in equation~(\ref{eq1:MPSquare}), we find an average of $\langle \mathcal{X}_{KS} \rangle = 5.38$. Because of experimental imperfections, the experimental violation of the Kochen-Specker
inequality falls short of the quantum-mechanical prediction. The dominating error source are imperfect unitary operations, in particular the entangling
gates applied up to six times in a single experimental run.

All experimental tests of hidden-variable theories are subject
to various possible loopholes. In our experiment, the
detection loophole does not play a role, as the state of the
ions are detected with near-perfect efficiency. From the
point of view of a hidden variable theory, still objections
can be made: In the experiment, the observables are not perfectly
compatible and since the observables are measured sequentially,
it may be that the hidden variables are disturbed during the sequence of
measurements, weakening the demand to assign to any observable
a fixed value independently of the context.

Nevertheless, it is possible to derive inequalities for classical
non-contextual models, wherein the hidden variables are
disturbed during the measurement process (see Methods).
More specifically, it can be proved that then the probabilities
of measurement outcomes obey the inequality
\begin{eqnarray}
\langle \mathcal{X}_{DHV} \rangle
&=&
\langle A_{12} A_{13}\rangle +
\langle A_{22} A_{23}\rangle +
\langle A_{12} A_{22}\rangle -
\langle A_{13} A_{23}\rangle
\nonumber
\\
&&
\!\!\!- 2 p^{\rm err}[A_{13}A_{12} A_{13}]
\!\!\!- 2 p^{\rm err}[A_{23}A_{22} A_{23}]
\label{chshnoise}
\\
&&
\!\!\!- 2 p^{\rm err}[A_{22}A_{12} A_{22}]
\!\!\!- 2 p^{\rm err}[A_{23}A_{13} A_{23}]
\leq 2.
\nonumber
\end{eqnarray}
Here,
$\langle A_i A_j \rangle = \langle v(A_i) v(A_j) \rangle$
denotes the ensemble average, if $A_i$ is measured before
$A_j$ and $p^{\rm err}[A_{i}A_{j} A_{i}]$ denotes the
probability that measuring $A_j$ introduces a change on the value
of $A_i$ if the sequence $A_{i}A_{j} A_{i}$ is measured.
To test this inequality, we prepared the state
$\Psi\propto |\!\uparrow\uparrow\rangle
+i\gamma|\!\downarrow\uparrow\rangle
+\gamma|\!\uparrow\downarrow\rangle
+{i|\!\downarrow\downarrow\rangle}$
where
$\gamma=\sqrt{2}-1$, and measured the Mermin-Peres square
with $\sigma_y$ and $\sigma_z$ exchanged. We find for the
whole square the value $\langle \mathcal{X}_{KS} \rangle =5.22(10)>4$, and for
Eq.~(\ref{chshnoise}) the value
$\langle \mathcal{X}_{DHV} \rangle = 2.23(5)>2$.
This proves that even disturbances of the hidden variables
for not perfectly compatible measurements cannot explain the
given experimental data.

In principle, our analysis of measurement disturbances and
dynamical hidden variable models can be extended to the
full Mermin-Peres square, however, the experimental techniques
have to be improved to find a violation there. Our findings show
already unambiguously that the experimentally observed phenomena cannot
be described by non-contextual models. Remarkably, here the experimental
observation of counter-intuitive quantum predictions did not require the
preparation of specific entangled states with non-local correlations.
We expect this result to stimulate new applications of quantum
contextuality for quantum information processing\cite{BP00, Galvao02, Nagata05,Spekkens09}.


\section*{Methods}
{\bf Quantum state detection.} The quantum state of a single ion is detected by illuminating both ions with light near the $S_{1/2}\leftrightarrow P_{1/2}$ transition frequency for $250\,\mu$s. To prevent the quantum information of the other ion from being read out, its $|\!\downarrow\rangle$ state population is transferred to (from) the $D_{5/2},m=5/2$ level before (after) the fluorescence measurement. The parameters of the read-out laser are set such that it keeps the axial COM-mode in the Lamb-Dicke regime\cite{Haffner08} with $\bar{n}_{COM}\approx 18$. By combining the counts of two photomultipliers, we observe a Poissonian distribution of photon counts in the detection window with average count numbers $n_\uparrow\approx 7.8$ and $n_\downarrow\approx 0.07$ in the bright and dark state, respectively. Setting the detection threshold to 1.5 counts, the conditional probabilities for wrong quantum state assignments amount to $p(\uparrow\!|\!\downarrow)\approx 0.24\%$ and $p(\downarrow\!|\!\uparrow)\approx 0.39\%$. At the end of the detection interval, the ion is optically pumped on the $S_{1/2}\leftrightarrow P_{1/2}$ to prevent leakage of population from the qubit level $|\!\downarrow\rangle$ to the state $|S_{1/2},m=-1/2\rangle$.\newline

{\bf QND measurements of spin correlations.} Quantum non-demolition measurements of observables $A_{ij}$ that measure spin correlations are carried out by mapping the subspace ${\cal H}_{A_{ij}}^+=\{\psi|A_{ij}\psi=\psi\}$ onto the subspace ${\cal H}_{\sigma_z^{(2)}}^+=\{\psi|\sigma_z^{(2)}\psi=\psi\}$ prior to the fluorescence measurement of $\sigma_z^{(2)}$. This is achieved by applying a unitary state transformation $U_{ij}$ satisfying $A_{ij}=U_{ij}^\dagger\sigma_z^{(2)}U_{ij}$ to the two-qubit state of interest. To decompose $U_{ij}$ into the elementary gate operations available in our setup, we use a gradient-ascent based numerical search routine\cite{Nebendahl09}. After measurement of $\sigma_z^{(2)}$, the inverse operation $U_{ij}^\dagger$ completes the QND measurement.\newline

{\bf Modeling imperfect measurements.} To deal with the case of imperfect measurements from a
hidden variable viewpoint, let us assume that there
is a hidden variable $\lambda$ which simultaneously
determines the probabilities of the results of all
sequences of measurements. Such probabilities are
written as $p[A^{(1)+},B^{(2)-}; AB]$ etc., denoting
the probability for the result $A^{(1)}=+1$ and
$B^{(2)}=-1$ when measuring first $A$ and then $B$.
Then the probabilities fulfill
\begin{eqnarray}
p[(A^{(1)+};A) \wedge (B^{(1)+};B)]
&\leq&
p[A^{(1)+}, B^{(2)+}; AB]\phantom{xxxx}
\nonumber
\\
&&
\hspace{-1.3cm}+
p[(B^{(1)+}; B) \wedge (B^{(2)-}; AB)].
\label{bound1}
\end{eqnarray}
This inequality holds because if $\lambda$ is such
that it contributes to $p[(A^{(1)+}; A) \wedge (B^{(1)+}; B)],$
then either the value of $B$ stays the same when measuring
$A$ and $\lambda$ contributes to $p[A^{(1)+}, B^{(2)+}; AB],$
or it is flipped and it contributes to
$p[(B^{(1)+}; B) \wedge (B^{(2)-}, AB)].$
The term $p[A^{(1)+}, B^{(2)+}; AB]$ is directly
measurable and we estimate the other term by assuming
\begin{equation}
p[(B^{(1)+}; B) \wedge (B^{(2)-}; AB)]
\leq
p[B^{(1)+},B^{(3)-}; BAB].
\label{bound2}
\end{equation}
This inequality means that the disturbance of
a predetermined value of $B$ caused by the measurement
of $B$ and $A$ should be larger than the disturbance
due to measurement of $A$ alone, as the former includes
additional experimental procedures compared with the latter.
The probability
$p[B^{(1)+},B^{(3)-}; BAB]$ is experimentally accessible
and combining Eq.~(\ref{bound1}) and (\ref{bound2})
we obtain a measurable upper bound on
$p[(A^{(1)+}; A) \wedge (B^{(1)+}; B)].$
Then, starting from the Clauser-Horne-Shimony-Holt-type
inequality\cite{Cabello08}
$\langle AB \rangle
+
\langle CD \rangle
+
\langle AC \rangle
-
\langle BD\rangle
\leq 2
$
which holds for simultaneous or undisturbed measurements,
using the above bounds and the notation
$p^{\rm err}[B A B]
= p[B^{(1)+},B^{(3)-}; BAB]+p[B^{(1)-},B^{(3)+}; BAB]
$,
one arrives at the inequality~(\ref{chshnoise}),
which then holds for sequences of measurements.

\section*{Acknowledgements}
We gratefully acknowledge support by the Austrian Science Fund (FWF), by the
European Commission (SCALA, OLAQUI and QICS networks + Marie-Curie program), by the Institut
f\"ur Quanteninformation GmbH, by the Spanish MCI Project No.
FIS2008-05596 and the Junta de Andaluc\'{\i}a Excellence Project No.
P06-FQM-02243. This material is based upon work supported in part by
IARPA.


\end{document}